\documentclass[twocolumn,english,prl,superscriptaddress]{revtex4}
\usepackage[T1]{fontenc}
\usepackage[latin9]{inputenc}
\usepackage{graphicx}

\usepackage{babel}
\usepackage{amsmath}
\usepackage{amsfonts}
\usepackage{times}

\begin{document}

\title{Hidden Noise Structure and Random Matrix Models of Stock Correlations}

\author{Ivailo I. Dimov}
\email{iid201@cims.nyu.edu}
\author{Petter N. Kolm}
\email{kolm@cims.nyu.edu}
\author{Lee Maclin}
\email{lee.maclin@gmail.com}
\author{Dan Y. C. Shiber}
\email{dcs255@cims.nyu.edu}
\address{Courant Institute of Mathematical Sciences, New York University\\ Corresponding author: Dan Y. C. Shiber}

\begin{abstract}
We find a novel correlation structure in the residual noise of stock
market returns that is remarkably linked to the composition and stability
of the top few significant factors driving the returns, and moreover
indicates that the noise band is composed of multiple subbands that
do not fully mix. Our findings allow us to construct effective generalized
random matrix theory market models \cite{LalouxEtal00_RMT,PlerouEtal_PRE02}
that are closely related to correlation and eigenvector clustering
\cite{Mantegna_HierarchicalClust,GopiEtal_ClustSectors}. We show
how to use these models in a simulation that incorporates heavy tails.
Finally, we demonstrate how a subtle purely stationary risk estimation
bias can arise in the conventional cleaning prescription \cite{LalouxEtal00_RMT}.
\end{abstract}
\maketitle

\paragraph{\underbar{Introduction:}}

Originally started in the context of nuclear physics \cite{Mehta_RMTbook},
random matrix theory (RMT) has thereafter found numerous applications
in a variety of fields such as number theory, disordered systems,
neural networks, and signal processing \cite{Mehta_RMTbook,SenguptaMitra99}.
Recently the pioneering work of Laloux \emph{et al} \cite{LalouxEtal00_RMT},
as well as much subsequent research \cite{PlerouEtal_PRE02,RMT_other},
have shown that RMT can also be a valuable tool for analyzing stock
market correlations, where noise can account for more than 2/3 of
the eigenvalue spectrum, and a typical large portfolio has size comparable
to the measurement time frame. Thus, much of the empirical eigenvalues
are spurious and represent measurement noise and biases. The remarkable
insight provided by Laloux \emph{et al} was to show that a suitable
fit to RMT can clean these spurious contributions, and moreover identify
the statistically significant signal, or common market risk factors
that drive the individual stock returns. The most prominent such non-idiosyncratic
factor is the nearly equal-weight top eigenvector, whose eigenvalue
is more than 20 times bigger than the average spectrum. Secondary
factors, are long-short portfolios of certain liquidity \cite{LalouxEtal00_RMT}
and industry structure \cite{GopiEtal_ClustSectors,AvellanedaLee_StatArb}, but their contribution
is typically an order of magnitude smaller. Most of the rest of the
eigenvectors are unstable in time, appear random, and their spectral
contribution can be fitted to the Marcenko-Pasteur (MP) distribution
\cite{MarcenkoPasteur} derived in the context of Gaussian RMT (GRMT).
The noisy eigenvalue correlations \cite{PlerouEtal_PRE02} also agree
with theory \cite{Mehta_RMTbook}. These results have been verified
over many stock selections, as well as return frequencies \cite{LalouxEtal00_RMT,PlerouEtal_PRE02,RMT_other}.

Despite the apparent success of the theory, subsequent research suggests
several empirical aspects that the original RMT cleaning may not account
for properly. (1) Tails and their correlations have non-trivial effects,
and are known to both broaden the spectrum above the upper noise-band
edge, as well sharpen it near the lower edge \cite{PlerouEtal_PRE02,BouchaudEtal_FatTailsRMT,RMT_fatTailsOther},
thus making the fit to the MP distribution problematic. The above
redistribution of spectral weight appears in conjunction with an enhancement
of the inverse participation ratios around both ends of the noise
spectrum, the so-called localization effect \cite{PlerouEtal_PRE02},
unlike GRMT where the participations are flat \cite{Mehta_RMTbook}.
(2) In addition to being partially localized, the band itself may
be split due to the same separation of correlation scales \cite{LilloMantegna}
that is thought to give rise to clustering of stocks between industries
\cite{Mantegna_HierarchicalClust}. So far this effect has not been
observed, however, due to the large amount of mixing that depletes
the stability of all the noisy eigenvectors. It is important to empirically
distinguish between the single and multiple band cases. (3) Non-stationarity
effects are insufficiently understood. They are suggested \cite{LalouxEtal00_RMT,PlerouEtal_PRE02}
to be the source of a residual bias in the risk estimates obtained
after RMT cleaning. However, in light of the abovementioned considerations,
it is not clear that the original cleaning procedures are unbiased
to begin with. 

In this work, we consider both $N=484$ $2$ minute S\&P500 TAQ midquote
returns between June 20 - Sep 20, 2007, as well as $N=451$ daily
S\&P500 returns between Jan 2001-Dec 2007 \cite{Foot_AutocorrRemoval}.
(1) We reveal a novel correlation structure of the residuals that
is linked to the structure and stability of the top few empirical
factors. Mainly, we find that the inverse participations of the localized
edge-eigenmodes of the band are dominated by the outlier stocks in
the composition of the top few factors, thus indicating that most
of the noise there is due to these stocks. The upper edge fluctuations
are mainly due to weakly correlated stocks with the smallest relative
weight in the market portfolio while lower edge fluctuations are due
to strongly correlated stocks identified as the outliers in the secondary
factors. The groups in the lower edge belong to major industrial sectors
\cite{PlerouEtal_PRE02,Mantegna_HierarchicalClust}, while the upper
edge contains a large diversified portfolio of medium to small liquidity
stocks. Moreover, because we find these groups to be disjoint, we
conclude that \emph{as long as the top few factors are stable and
distinct, the noise band is composed of multiple subbands that do
not fully mix.} (2) We pinpoint the effective positive-definite cleaned
matrices that exhibit the multi-residual and factor structure above
to be the hierarchical RMT models \cite{LilloMantegna}, which are
closely related to coarse-grained {}``real space'' models of market
clustering \cite{Mantegna_HierarchicalClust}, and fundamentally arise
out of correlation scale separation. (3) We use these effective models
to perform a one-factor stochastic volatility \cite{BouchaudEtal_FatTailsRMT}
simulation in order to take into account the effect of tails and their
correlations. (4) We show how conventional cleaning can give rise
to a subtle purely stationary risk-estimation bias.

\begin{figure}
\includegraphics[width=8.5cm,height=7cm]{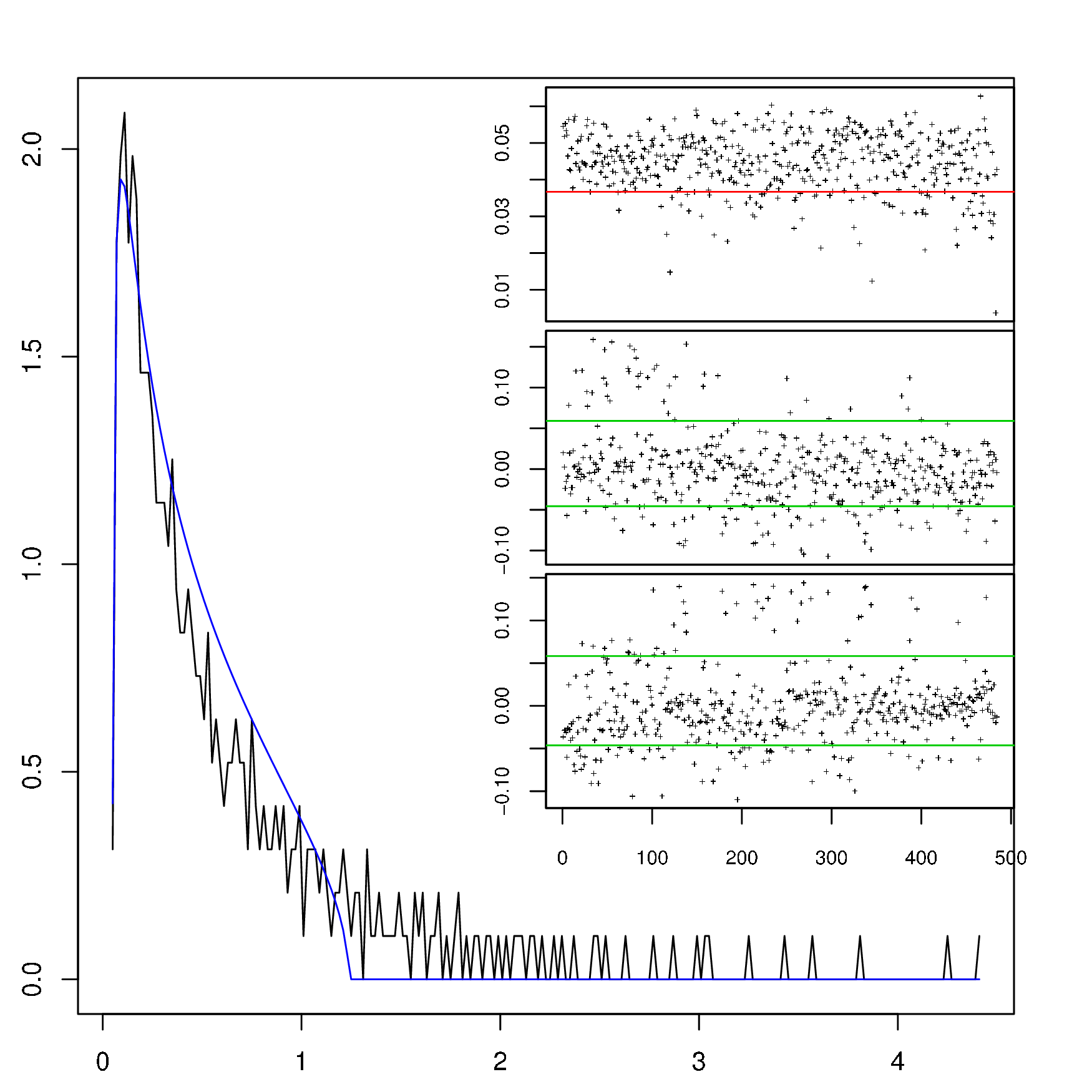} 

\caption{\emph{Main}: A fit of the MP distribution to $2$ min data for $N=484,$
stocks in the S\&P 500 with $T=3N$ yields $Q_{eff}=2.25$, and $\sigma_{eff}=0.67$.
The top three eigenvalues, $\lambda_{1}=152.9$, $\lambda_{2}=8.2$,
$\lambda_{3}=7.6$, $\lambda_{4}=5.3$, $\lambda_{5}=5.2$ were omitted
from the plot due to their scale. \emph{Inset}: The top three eigenvectors,
$e_{ki},$ $k=1,2,3$, with their entries $i$ sorted by decreasing
liquidity (from left to right). Note the significant outliers in each
$e_{k}$ as emphasized by the horizontal lines.}

\end{figure}

\paragraph{\underbar{Empirical Results:}}

Given our set of $N$ stock series $S_{i}(t)$, $t=1,\ldots,T$, from
their log returns $x_{t,i}=\log(S_{t,i}/S_{t-1,i})$ we calculate
the empirical correlation $C_{E}=(\langle x_{i}x_{j}\rangle-\langle x_{i}\rangle\langle x_{j}\rangle)/\sigma_{i}\sigma_{j}$
where $\sigma_{i}=\sqrt{\langle x_{i}^{2}\rangle-\langle x_{i}\rangle^{2}}$.
According to RMT \cite{Mehta_RMTbook,SenguptaMitra99,LalouxEtal00_RMT},
if $C_{E}$ were obtained from a purely random signal of bounded variance
$\sigma$ whose marginal tails are not too heavy \cite{BouchaudEtal_FatTailsRMT},
then in the limit $N\rightarrow\infty$ with $Q\equiv N/T$ fixed,
the correlations will self-average and will have an asymptotically
deterministic eigenvalue spectrum given by the MP distribution \cite{MarcenkoPasteur}:

\begin{equation}
p_{MP}(\lambda)=\frac{Q}{2\pi\sigma^{2}}\frac{\sqrt{(\lambda_{+}-\lambda)(\lambda-\lambda_{-})}}{\lambda}.\label{eq:MP}\end{equation}
In the above, the eigenvalues $\lambda$ are restricted to lie within
the hard-edge spectral band, $\lambda\in[\lambda_{-},\lambda_{+}],$
with $\lambda_{\pm}=\lambda_{\pm}(Q,\sigma)=\sigma^{2}(1\pm\sqrt{1/Q})^{2}$.
One can interpret (\ref{eq:MP}) as the the finite $T/N$ noise-induced
broadening and bias away from the underlying trivial spectrum $p_{clean}(\lambda)=\delta(\lambda-1)$
that $p_{MP}(\lambda)$ reduces to in the limit $Q\rightarrow\infty$. 

Of course, stock market correlations are not purely random, so a fit
for $\sigma$ and $Q$ is necessary in (\ref{eq:Ceff}) if one wants
to identify the trully residual part of the spectrum \cite{LalouxEtal00_RMT,PlerouEtal_PRE02}.
In Fig. 1 we show such a fit to the noisy region of the $2$ min data
that yields $\sigma_{eff}=0.67,$ $Q_{eff}=2.25.$ Note that much
of the spectrum lies outside the MP band. In the inset of Fig 1 we
plot the composition, of the top three eigenvectors $\{e_{k}\}$,
$k=1,2,3$ sorted by decreasing liquidity. Unlike the prediction of
GRMT where $e_{k}$ should be a mean-zero unit Gaussian, there are
clear deviations from such behavior in all three eigenvectors, as
emphasized in the inset. In fact, $e_{1}$ has non-zero mean, $\langle e_{1i}\rangle_{i}=0.044\simeq N^{-1/2}$,
representing a long-only market portfolio, while $e_{2}$ and $e_{3}$
represent long-short portfolios. All three $e_{k}$ can be interpreted
as significant common factors. Furthermore, as is clear from the plot,
the outliers in the factor composition have certain liquidity structure.
In the case of $e_{2}$ and $e_{3}$ in of Fig 1, these outliers can
be identified with major sectors such as financials, oil, and utilities,
whose correlations are relatively stable in time \cite{GopiEtal_ClustSectors,AvellanedaLee_StatArb,Mantegna_HierarchicalClust}. 

Despite the appearance of factors, one expects the random residual
spectral contribution to be well fitted to RMT. However, there are
important issues with the fit to $p_{MP}$ that one needs to address.
Heavy tails in the multivariate distribution of $x_{t,i}$ have non-trivial
effects. They are known to broaden the spectral weight above the upper
edge, as well as sharpen it near the lower edge \cite{PlerouEtal_PRE02,BouchaudEtal_FatTailsRMT,RMT_fatTailsOther},
both features readily noticeable in Fig. 1 as well as in daily data
\cite{LalouxEtal00_RMT}. Moreover, such tails tend to induce outliers
in the composition of the principal components near the band edge
\cite{BouchaudEtal_FatTailsRMT} causing deviations from the standard
Gaussian distribution of the composition expected by GRMT and inducing
localization. Indeed, just as in daily data \cite{PlerouEtal_PRE02}
we see in the $2$ min returns that the eigenvectors are localized
at both ends of the noise spectrum by computing the \emph{inverse
participation ratio} \cite{PlerouEtal_PRE02}, $I_{k}=\sum_{i=1}^{N}[e_{ki}]^{4}$
for each eigenvector $e_{k}$. Intuitively, the \emph{participation}
\emph{$P_{k}\equiv1/I_{k}$} scales as the number of non-trivial entries
in a normalized $e_{k}$: $P_{k}=N$ for equal weight vectors, while
$P_{k}=1$ for a single non-trivial weight. As evident in Fig 2 (a),
the participation is strongly localized near the band edges indicating
that the eigenvectors there are dominated by outliers.

A nice trick that avoids estimating the effects of tails is to clean
the noisy eigenvalues $\Lambda_{E}=diag(\lambda_{1},\ldots,\lambda_{K};\{\lambda_{noise}\})$
of $C_{E}=S_{E}\Lambda_{E}S_{E}'$ with a flat band with scale proportional
to $\sigma_{eff}$ while working in the original eigenvector basis
$S_{E}=(e_{1},\ldots,e_{N})$ \cite{LalouxEtal00_RMT}, thus obtaining
a {}``filtered'' matrix \cite{PlerouEtal_PRE02}. In fact, unless
the empirically measured tail fluctuations significantly break the
rotational invariance implied by cleaning with a flat band, $\sigma_{eff}$
can be thought of as the overall scale of the residual noise that
one can tune even without fitting to the Gaussian formula. The feasibility
of this {}``filtering'' procedure can also be justified with the
resulting significant improvement of the portfolio risk estimates
that one obtains with cleaning \cite{LalouxEtal00_RMT,GopiEtal_ClustSectors}. 

\begin{figure}
\includegraphics[width=8cm,height=6cm]{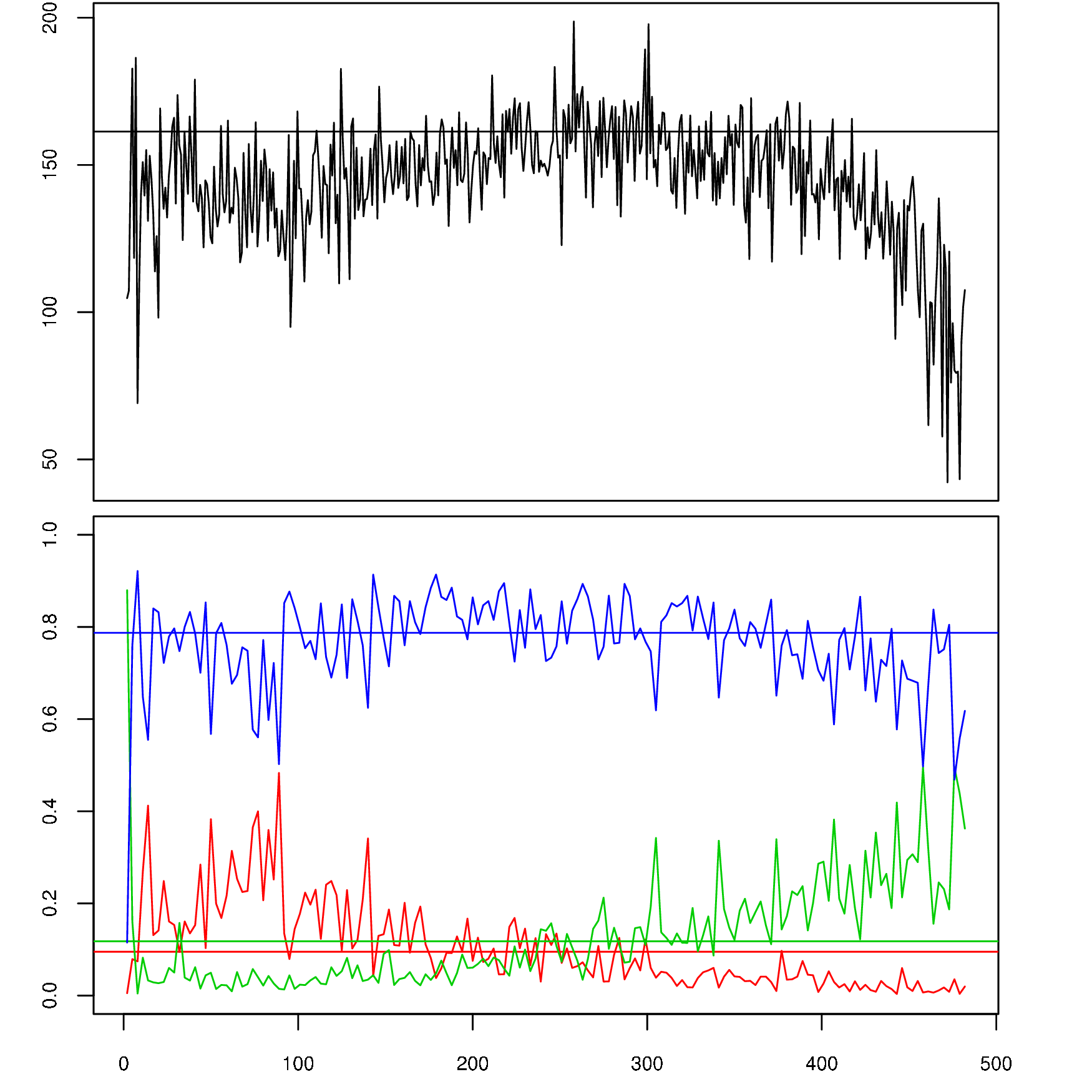}

\caption{\emph{Top}: The participation $P_{k}$ of the eigenvectors $e_{k}$,
$k\geq2$ for the data in Fig 1 exhibits localization. Flat horizontal
line represents $P_{k}^{GMRT}=N/3.$ \cite{Mehta_RMTbook}.\emph{
Bottom}: Relative inverse participation $R_{k}^{(G)}$ for the groups
$G_{1}$ (red), $G_{23}$ (green) and $G^{\perp}$ (blue) defined
in the text. Flat lines represent $R_{k,RMT}^{(G)}$.}

\end{figure}

We will now show, however, that because of a novel structure of the
eigenvectors, cleaning with a flat band $\sigma_{eff}$ is inconsistent
with their symmetry. Suppose we look at the following group of stocks,
$\{G_{1},G_{k}^{\pm},G^{\perp}\},$ $k=2,3$ selected so that $G_{k}^{\pm}$
contains the top/bottom outliers $|e_{ki}|>\tilde{e}_{k}$ of $e_{k}$
above/below a certain threshold $\tilde{e}_{k}$ (see Fig 1 inset),
$G_{1}$ are the outliers of smallest absolute weight in $e_{1}$,
and $G^{\perp}$ are all the other stocks. We find that for reasonable
threshold values, $\tilde{e}_{k}\simeq1.5\sigma_{e_{k}}$, \emph{$G_{1}$
}is \emph{disjoint} from\emph{ $G_{23}\equiv\bigcup G_{k}^{\pm}.$
Moreover, the relative contribution of each group G to the inverse
participation,}

\begin{equation}
R_{k}^{(G)}=\frac{\sum_{i\in G}[e_{ki}]^{4}}{I_{k}}\equiv\frac{I_{k}^{(G)}}{I_{k}},\label{eq:Rel_Ik}\end{equation}
\emph{is inhomogeneously distributed across the noise band as shown
in Fig 2 (b), }so that $G_{1}$ contributes mostly to the upper edge
while $G_{23}$ contributes mostly to the lower edge. Because this
behavior is inconsistent with homogeneous cleaning, we interpret it
as an indication that \emph{the noise band is composed of multiple
subbands that do not mix. }In particular, the three groups above form
a partition of all the stocks, where the number of assets in each
subgroup, or the \emph{group degeneracies,} for the data in Fig 1
are $\{D_{1},D_{23},D^{\perp}\}=\{46,61,377\}$.

\paragraph{\underbar{Constructing Coarse-Grained Effective Models:}}

The partition above is reminiscent of partitions previously obtained
by {}``real space'' hierarchical clustering of stocks into industries
\cite{Mantegna_HierarchicalClust}, which can be thought of as arising
from a coarse-grained separation of correlation scales, also known
to give rise to multiple subbands in the spectrum \cite{LilloMantegna}.
Moreover, it has been observed \cite{GopiEtal_ClustSectors} that
clustering of significant \emph{eigenvector} \emph{components} results
in the same industries as those obtained from the real space procedure,
and arises out of a mean-field duality relation between the two approaches
\cite{DimovShiber_unpublished}. \emph{Therefore, we interpret the
multiple subband structure above as arising from a particular type
of underlying separation of correlation scales apparent at the time
scales of measurement of $C_{E}$. }We have observed the above multiple
band structure in both $2$ min and daily data.

\begin{figure}
\includegraphics[width=8cm,height=6cm]{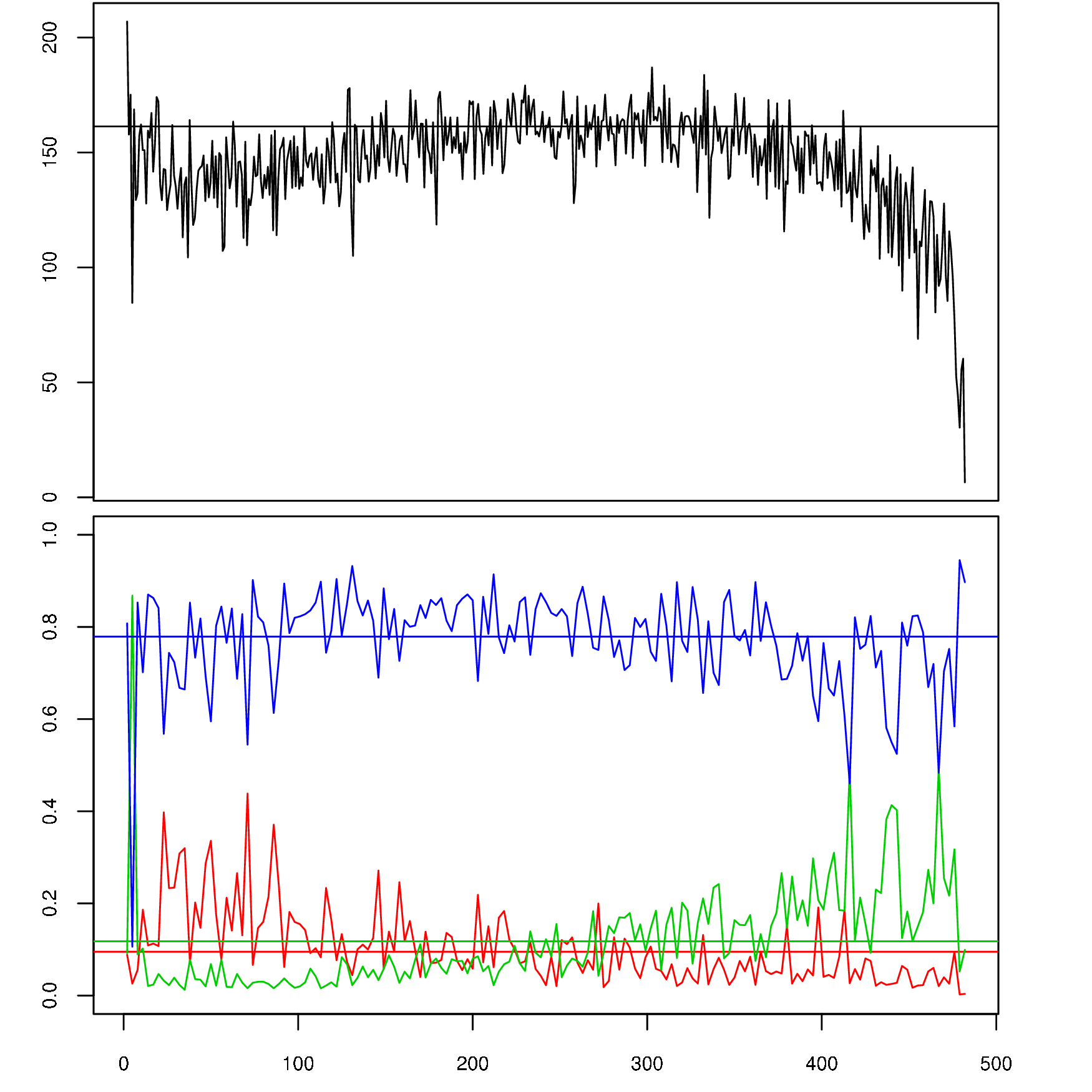}

\caption{The same quantities as in Fig 2, except all the data was \emph{simulated}
from the effective model (\ref{eq:Ceff}) by taking into account tail
effects in the multivariate distribution via the one-factor stochastic
volatility model \cite{BouchaudEtal_FatTailsRMT} with tail index
$\nu=3$. }

\end{figure}

Let us gain insight into the details of this correlation structure
for the case of the data in Fig 1. By clustering analysis \cite{Mantegna_HierarchicalClust}
we find that $G_{23}$ separates further, $G_{23}=\{G_{23}^{+},G_{23}^{-}\}$,
into two nearly-equally large groups of distinctly higher/lower mean
average correlation with degeneracies $\{D_{23}^{+},D_{23}^{-}\}=\{29,32\}$
respectively. For this sample, we find that $G_{23}^{+}$ contains
Electric Utilities, as well as Oil \& Gas Drilling \& Exploration
stocks, while $G_{23}^{-}$ contains Oil \& Gas Exploration, as well
as some Financial stocks. From the average correlation between all
four groups, we thus construct the following {}``minimal'' coarse-grained
\emph{effective model}:

\begin{equation}
C_{eff}^{N\times N}=\left(\begin{array}{cccc}
C_{23+}^{D_{23}^{+}\times D_{23}^{+}} & 0.36 & 0.35 & 0.22\\
0.36 & C_{23-}^{D_{23}^{-}\times D_{23}^{-}} & 0.33 & 0.20\\
0.35 & 0.33 & C_{\perp}^{D^{\perp}\times D^{\perp}} & 0.20\\
0.22 & 0.20 & 0.20 & C_{1}^{D_{1}\times D_{1}}\end{array}\right)\label{eq:Ceff}\end{equation}
which can be readily checked to be positive definite. The entries
of each $D\times D$ diagonal block above are $C^{D\times D}=(1-\rho_{G})I_{D}+\rho_{G}^{D\times D},$
with $\{\rho_{23}^{+},\rho_{23}^{-},\rho_{\perp},\rho_{1}\}=\{0.59,0.48,0.32,0.13\}$
respectively being the average correlation of each of the four groups
$G$ and $\rho_{G}^{D\times D}$ is a block whose entries are $\rho_{G}$.
One can check that (\ref{eq:Ceff}) also gives rise to $4$ distinct
factors and $4$ distinct subbands. 

Note that unlike the strongly correlated ones in $G_{23}$, the stocks
in $G_{1}$ are typically not easily detectable with conventional
hierarchical clustering approaches \cite{Mantegna_HierarchicalClust},
although they are distinctly visible if one looks at the top factor
(see Fig 1 inset). Indeed, being weakly correlated between each other
as well as with the rest of the market, these stocks will not appear
in localized real-space clusters but instead will group with other
stocks in later stages of the hierarchy. At the same time, both the
degeneracy $D_{1}$ and overal risk contribution of $G_{1}$ are comparable
to those of the localized sectors, as also directly suggested by Fig
2 (b). To properly account for the separation of correlation scales
in markets, one must also include the contribution of the weakly correlated
stocks.

\paragraph{\underbar{Simulating with tails:}}

A check of the validity of the effective model (\ref{eq:Ceff}) is
ultimately provided if one can reproduce the empirical spectrum and
participations through simulation. To do so, one must properly take
into account heavy tailed behavior of actual returns. It is known
that such tails can be induced by heteroskedasticities of the underlying
stock volatilities \cite{BouchaudPotters_book}, although the details
of the correlations of such volatility dynamics are not well understood.
We thus use the simplest multivariate conditional Gaussian model with
one-factor variance-gamma volatilities, which is known to produce
a Student-t type of series \cite{BouchaudEtal_FatTailsRMT} for the
joint returns. We use an inverse-gamma tail index of $\nu\simeq3$.
The resulting spectrum \cite{BouchaudEtal_FatTailsRMT,RMT_fatTailsOther}
agrees well with the empirical one. Moreover, comparing Figs 2 and
3, we see that the inverse participations are also in good agreement.

\paragraph{\underbar{A Subtle Stationary Bias:}}

The discussion so far suggests that even for stationary data, RMT
cleaning could produce biased risk estimates. Let us demonstrate this
for the simplest case of multivariate Gaussian returns simulated with
the effective model (\ref{eq:Ceff}). Without loss of generality we
normalize the returns to mean zero unit variance. Using the notation
in \cite{PlerouEtal_PRE02}, the predicted risk of a portfolio $w=(w_{1},\ldots,w_{N})$
is $\Omega_{p}^{2}=w\cdot C_{clean}\cdot w^{T}$. The portfolios we
look at are equal-weight average representatives of different subbands
$K$, $w_{K}=\sum_{k\in K}e_{k}$, where $\{e_{k}\}$ are the eigenvectors
of the \emph{effective model} (\ref{eq:Ceff}). Moreover, instead
of a {}``budget constraint'' \cite{PlerouEtal_PRE02}, we impose
a {}``risk constraint'' by normalizing $w_{K}$ to unit norm. We
then compute at every forecasting period the relative difference between
realized and predicted risk, $\delta r=(\Omega_{r}^{2}-\Omega_{p}^{2})/\Omega_{p}^{2}.$
For the subbands $\{K_{1},K^{\perp},K_{23}\}$ corresponding to the
groups of stocks that enter in Fig 2 (b), we find respective biases
$\delta r_{i}=\{26\pm4\%,2\pm4\%,-17\pm2\%\}.$ Note that although
$\delta r_{1}$ and $\delta r_{23}$ are significant, they are of
opposite sign. Indeed, we have checked that all three contributions
nearly cancel when one looks at the relative realized versus predicted
risk of the entire noise band, $\delta r_{all}=2\pm3\%$. Finally,
we also observe significant biases $\delta r_{1}$ and $\delta r_{23}$
in the actual data. However, in this case, there are subtleties in
disentangling the effects of multiple bands, tails, and non-stationarity.
We postpone discussing these effects, as well as multi-residual generalizations
of the RMT cleaning procedure to later work \cite{DimovShiber_unpublished}.

\paragraph{\underbar{Summary: }}

In conclusion, we have found strong evidence that instead of homogeneous,
the stock market correlation residuals are composed of multiple subbands
that do not fully mix. This structure is manifested through an asymmetry
in the relative inverse participations of the eigenvectors within
the noise band, which is inconsistent with purely symmetric cleaning
that doesn't distinguish between different parts of the noise spectrum.
The multi-residual picture above natrually emerges from market models
with multiple correlation scales, that we have identified and simulated.
As a direct consequence, the scale separation within the noise band
also produces inhomogeneities in the effective residual risk that
in turn induce purely stationary biases of the original RMT cleaning.

\begin{acknowledgements}
We would like to thank Marco Avellaneda and Jim Gatheral for their insightful comments and discussion.
\end{acknowledgements}


\begin{thebibliography}{16}
\bibitem{Foot_AutocorrRemoval} We have performed all the necessary
tests \cite{BouchaudPotters_book} to ensure that the marginal $2$
min midquote returns have no residual autocorrelations.
\end{thebibliography}
\end{document}